\documentclass[aps,prl,twocolumn,nofootinbib,groupedaddress,amsfonts,floatfix]{revtex4}

\usepackage{graphicx,amsmath,amssymb,amstext}
\usepackage{amssymb,amsbsy,amsfonts,amsthm,color}

\newcommand{\gsim}{\;\mbox{\raisebox{-0.5ex}{$\stackrel{>}{\scriptstyle{\sim}}$}}\;}
\newcommand{\Beq}{\begin{equation}\begin{aligned}}
\newcommand{\Eeq}{\end{aligned}\end{equation}}
\newcommand{\mpl}{M_{\textrm{pl}}}

\begin{document}

\title{Oscillons After Inflation}

\author{Mustafa A. Amin${}^1$}
\author{Richard Easther${}^2$}
\author{Hal Finkel${}^2$}
\author{Raphael Flauger${}^2$}
\author{Mark P. Hertzberg${}^3$}

\affiliation{${}^1$Department of Physics, Massachusetts Institute of Technology, Cambridge MA 02139}
\affiliation{${}^2$Department of Physics, Yale University, New Haven CT 06520}
\affiliation{${}^3$SITP and KIPAC, Department of Physics, Stanford University, Stanford CA 94305}
 
\begin{abstract}
Oscillons are massive, long-lived, localized excitations of a  scalar field.  We show that in a class of well-motivated single-field models, inflation is followed by self-resonance, leading to copious oscillon generation and a lengthy period of oscillon domination. These models are characterized by an inflaton potential which has a quadratic minimum and is shallower than quadratic away from the minimum.  This set includes both string monodromy models and a class of supergravity inspired scenarios, and is in good agreement with the current central values of the  concordance cosmology parameters. We assume that the inflaton is weakly coupled to other fields, so as not to quickly drain energy from the oscillons or prevent them from forming.  An oscillon-dominated universe has a greatly enhanced primordial power spectrum on very small scales relative to that seen with a quadratic potential, possibly leading to novel gravitational effects in the early universe.
\end{abstract}

\maketitle
 \noindent
 
 Simple, single-field models of inflationary cosmology are often associated with energy scales far beyond the reach of present day accelerators and the properties of the post-inflationary universe are largely unknown. One constraint on this phase is  that energy must be extracted from the oscillating inflaton condensate, ensuring that the universe becomes radiation dominated,  setting the scene for the hot big bang, and the production of  the cosmological neutrino background and  nucleosynthesis.  A widely-studied candidate for this process is parametric resonance \cite{Traschen:1990sw,Shtanov:1994ce,Kofman:1994rk}. In many cases, the potential can be {\em self}-resonant, where resonance generates quanta of the inflaton field itself more efficiently than particles  coupled to the inflaton. We show that in a  class of well-motivated self-resonant models, the universe may become dominated by oscillons: massive, localized, metastable configurations of a scalar field \cite{Bogolyubsky:1976yu,Gleiser:1993pt,Copeland:1995fq, Broadhead:2005hn, Gleiser:2009ys,Gleiser:2011xj, Amin:2010jq,Amin:2010xe,Gleiser:2010qt, Amin:2010dc}.

In this letter, we study a single inflaton, $\phi$, with a canonical kinetic term and  potential, $V(\phi)$, minimally coupled to Einstein gravity.  Oscillons can form if
\begin{equation} \label{eq:potU}
V(\phi)=\frac{m^2\phi^2}{2}+U(\phi),
\end{equation}
where $U(0)=0$ and $U(\phi)<0$ for some range of $\phi$ (see, e.g. \cite{Amin:2010jq}). Consider potentials with $V(\phi)\sim\phi^{2\alpha}$ during inflation and $\alpha<1$. These are generated by a number of  string and supergravity scenarios  \cite{Silverstein:2008sg,McAllister:2008hb,Flauger:2009ab,Dong:2010in,Kallosh:2010xz,Kallosh:2010ug,Dubovsky:2011tu}, and  yield $U(\phi)<0$ at large $\phi$. We require that $V(\phi)$ has a stable minimum which we chose to be at the origin, so it is natural to expect that $V(\phi)\sim\phi^{2}$ for small $\phi$.  Finally, by continuity, there is necessarily some crossover scale, $\phi\approx M$, between these two regimes.  We capture this with the following explicit potential
 \begin{equation}
\label{eq:masterpot} V(\phi) =
   \frac{m^2M^2}{2\alpha} \left[ \left( 1 + \frac{\phi^2}{M^2} \right)^{\!\!\alpha} -1\right] \, .  
 \end{equation}
 The precise forms of $V(\phi)$   in  scenarios with $V\sim\phi^{2\alpha}$ can differ from equation~(\ref{eq:masterpot}): however, our results suggest that while oscillon formation is sensitive to  $M$, it is insensitive to the detailed form of the potential.   Moreover, for $\alpha=1/2$ we reproduce the axion monodromy potential \cite{McAllister:2008hb,Flauger:2009ab}. We stipulate that  the couplings between the inflaton and other fields are small enough for them to be ignored.
 
 The  tensor-scalar ratio, $r$, and scale-dependence in the scalar perturbations, $|n_s-1|$,   grows with $\alpha$ (see, e.g. \cite{Adshead:2010mc}). Quartic inflation $(\alpha=2)$ is  ruled out  by current data \cite{Peiris:2003ff,Komatsu:2010fb,Mortonson:2010er} and even quadratic inflation $(\alpha=1)$ is somewhat disfavored, relative to models with $\alpha < 1$ \cite{Martin:2010kz}.  Consequently, the above potential is well-motivated, both theoretically and phenomenologically.\footnote{Oscillon production in hybrid inflation models is studied in \cite{Gleiser:2011xj}. These models have $n_s\ge1$, and are disfavored by observations.}  

The post-inflationary universe is initially smooth, so even if a potential supports oscillon solutions,  an actual oscillon-dominated phase requires a mechanism for generating inhomogeneity within the post-inflationary horizon (see, e.g. \cite{Amin:2010xe}).  Equation~\ref{eq:masterpot}  supports parametric resonance when $\alpha<1$, which  lead to the explosive production of $\phi$ quanta, and a highly inhomogeneous universe. However,  $M$ is large, the $V(\phi)$ effectively quadratic during both the last portion of inflation and subsequent oscillatory phase, suppressing resonance and oscillon production.  Conversely, if $M$ is significantly sub-Planckian we see resonance and oscillons can form. Note that narrow resonance also occurs  when $\alpha>1$, but $V(\phi)$ cannot support oscillons.    
     
In what follows, we first summarize the inflationary dynamics and describe a  Floquet analysis of the resonant phase.   We show that strong resonance and a subsequent oscillon-dominated phase requires $0 \le \alpha \lesssim0.9$ and $M \lesssim 0.05 \mpl$ ($\mpl\equiv 1/\sqrt{8\pi G_N}$), which 
may be realized in the physical scenarios that motivate these models. 
We then discuss the cosmological consequences of an oscillon-dominated phase.

 \section{Inflationary Dynamics}
The observed amplitude of the primordial fluctuations effectively removes one free parameter from the potential in equation~\eqref{eq:masterpot}.   Further, we will see that we are primarily interested in models where $M$ is substantially smaller than the Planck mass, so that $V(\phi) \approx m^2 M^2 (\phi/M)^{2 \alpha}/{2\alpha}$ during inflation. 

Astrophysically interesting perturbations are laid down when the remaining number of e-folds before the end of inflation, $N \sim55$, though in general, $N$ is a function of the post-inflationary expansion history \cite{Adshead:2010mc}. Using standard slow-roll approximations, the amplitude of the power spectrum  of curvature perturbations is
\begin{equation}
\Delta_R^2 = \frac{1}{96 \pi^2  \alpha^3} \left(\frac{m}{\mpl}\right)^2    \left(  \frac{ M}{\mpl} \right)^{2-2\alpha }  (4\alpha N)^{1+\alpha}.
\end{equation}
For a given $\alpha$ and $\beta \equiv \mpl/M$, we  use the above equation with $N=55$ and $ \Delta_R^2 = 2.4 \times 10^{-9}$ \cite{Komatsu:2010fb} to deduce $m$. 

\section{Resonance and Oscillons}
 
Oscillon production at the end of inflation with $\qquad U(\phi)=-\lambda \phi^4/4 +g^2\phi^6/{6 m^2}$
%
%
was studied in \cite{Amin:2010xe, Amin:2010dc}. When $(\lambda/g)^2 \ll 1$, oscillons are  copiously generated, with properties matching analytic predictions  \cite{Amin:2010xe, Amin:2010jq}.  However,  an inflationary phase where $V(\phi)$ is dominated by a $\phi^6$ term has an unphysical perturbation spectrum. If the above $U(\phi)$ is viewed as a truncation of  eq.~(\ref{eq:masterpot}) then $(\lambda/g)^2 \sim 1$, and we cannot appeal to the results of  \cite{Amin:2010xe, Amin:2010jq, Amin:2010dc} for the properties of oscillons. Thus, to study
oscillon formation in this physically reasonable scenario, we rely on numerical simulations.

\begin{figure}[tb] 
   \centering
   \includegraphics[width=2.7in]{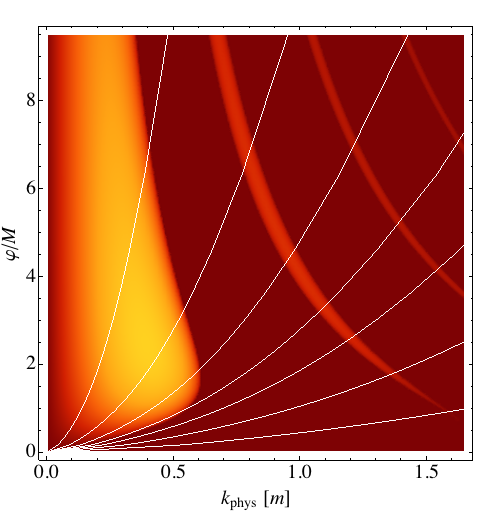} 
   \caption{Floquet diagram with $\alpha=1/2$, $\beta=100$. The stable regions are dark red. Within the unstable bands, lighter colors correspond to  larger real-valued Floquet exponents. White lines show $k/a(t)$ for representative  modes in an expanding universe.
}
   \label{fig:floq}
\end{figure}

We can gain a heuristic understanding of oscillon formation by looking at the instability diagram for the potential in \eqref{eq:masterpot}, as resonance generates large inhomogeneities which then relax to form oscillons \cite{Amin:2010xe}. Ignoring expansion and working in the limit where $\phi$ is approximately homogeneous, Floquet theory allows us to write the individual momentum modes of $\phi$  as
\begin{equation}
\phi_k =   P_+(t)e^{\mu_k t} +P_{-}(t)e^{-\mu_kt}
\end{equation} 
where $P_{\pm}(t)$ are periodic functions and $\pm\mu_k$ are called Floquet exponents. Our first task is to calculate these exponents:
  if the real part of $\mu_k$, $\Re(\mu_k)$, is nonzero and its magnitude is larger than the Hubble parameter, $H\sim t^{-1}$, at the end of inflation,  the mode will grow. Roughly speaking, if $|\Re(\mu_k)| /H \gsim 10$, we have strong resonance. In an expanding universe,  $\phi_k$ has a physical wavenumber $k/a(t)$ and thus moves through a number of Floquet bands  as the scale factor, $a(t)$, grows, as shown in  Figure~\ref{fig:floq}. For our potential, with  $\beta = \mpl/M$, one can show  that the maximum value of $|\Re(\mu_k)|/H$ as the modes traverse the Floquet bands is $[|\Re(\mu_k)|/H]_{max}\approx A(\alpha)\beta$ where $A(\alpha)\approx (1/2)\left[(1-\alpha)-(1/10)(1-\alpha)^2\right]$.

We studied the nonlinear dynamics of
 resonance following inflation driven by eq.~(\ref{eq:masterpot}) using {\sc PSpectRe} \cite{Easther:2010qz}.    {\sc PSpectRe} solves the fully nonlinear three dimensional Klein-Gordon equation in an expanding background whose behavior is governed by the usual Friedman equations, sourced by the average density and pressure. The backreaction of  metric perturbations on the field is ignored.  Our simulations begin at the first instant  $\dot{\phi}=0$, although our results are insensitive to the details of this choice. The scalefactor $a=1$ at the beginning of our simulations. We assume a standard spectrum of initial vacuum fluctuations, although we checked that our results are qualitatively insensitive to the detailed form of the initial conditions. We ignore backreaction of the metric perturbations on the field evolution -- these can be shown to be small during resonance. The initial  boxsize is $L=25/m$ with $256^3$  points in the (comoving) simulation volume. 

\begin{figure}[tb] 
   \centering
   \includegraphics[width=2.7in]{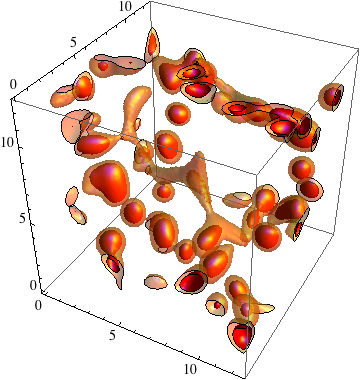}
    \includegraphics[width=2.8in]{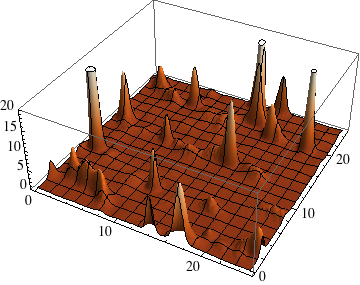}
   \caption{ Oscillon configuration  with $\alpha = 1/2$ and $\beta = 50$. The top plot shows regions where $\rho/\langle \rho\rangle >4 $ (transparent) and 12 (solid), while the lower plot shows $\rho/\langle \rho\rangle$ on a two dimensional slice through the simulation.  Length units are $1/a(t) m$, and these plots were made when $a(t)=5.46$. }
   \label{fig:visualize}
\end{figure} 

\begin{figure}[tb] 
   \centering
   \includegraphics[width=2.8in]{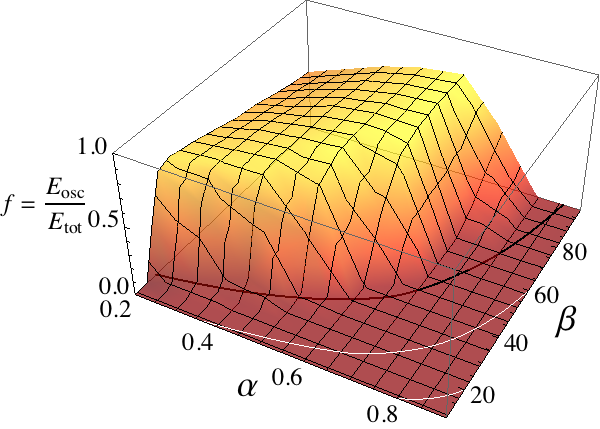} 
   
   \mbox{}
   
   \caption{ The    statistic, $f$, is shown at $a(t) =7$ ($a(t) =1$ at the beginning of the simulation) as a function of $\alpha$ and $\beta= M_{\rm{pl}}/M$.   Contours show maximal value of the $[|\Re(\mu_k)|/H]_{max}$. The thick black contour denotes $[|\Re(\mu_k)|/H]_{max}=7$ whereas the thin white ones correspond to  $[|\Re(\mu_k)|/H]_{max}=1,3$. }
   \label{fig:result}
\end{figure} 
 
A single timeslice of a representative simulation is shown in Figure~\ref{fig:visualize}.   Given that oscillons are large over-densities,  a {\em necessary\/} condition for oscillon domination is that
\begin{equation}
f = \frac{\int_{\rho> 2 \langle \rho\rangle} \rho dV }{ \int{ \rho dV}} \, ,
\end{equation}
 the fraction of the total energy density contributed by regions where $\rho/\langle \rho\rangle >2$, is nontrivial.  Oscillons are effectively fixed in space,  persisting for a Hubble time or more, so the overall density in an oscillon-dominated universe at different times is strongly correlated. Heuristically,   $f \gsim 0.3$ is {\em sufficient\/}  to guarantee that the field configuration (and thus the post-inflationary universe as a whole) was dominated by oscillons.   Figure~\ref{fig:result} shows $f$ as a function of $\alpha$ and $\beta$, along with  the maximal value of the resonance parameter $[|\Re(\mu_k)|/H]_{max}$. We see that strong resonance, or $[|\Re(\mu_k)|/H]_{max}\gtrsim 10$, is both necessary and sufficient for prompt, copious oscillon formation. 
  
In models for which $f$  is non-zero, it remains approximately constant for a several Hubble times after the onset of oscillon domination, demonstrating that this phase is long-lived, relative to prevailing cosmological time scales.  Unlike the oscillons studied in \cite{Amin:2010xe, Amin:2010jq, Amin:2010dc} which have a stable, radial envelope, $\Phi(r)$, which evolves very  slowly with time, here
 the corresponding envelope is  a periodic function of time, and the oscillon ``breathes'' in and out. 
The detailed dynamics of these oscillon solutions will be discussed in a future publication, but  we have simulated a single oscillon (ignoring expansion) over a long interval for representative values of $\alpha$ and $\beta$ after imposing strict radial symmetry, reducing the problem to a 1+1 PDE.  Even though oscillons are not protected by a conserved charge and radiate energy \cite{Segur:1987mg,Fodor:2008du, Hertzberg:2010yz}, these simulations suggest that  they live long enough for the universe to grow by a factor of 100 or more, and we expect this to be true even if the assumption of radial symmetry is dropped. Also, the quantum radiation will be small in the regime where the self couplings, such as $\lambda\sim m^2/M^2$, are small \cite{Hertzberg:2010yz}.  
 
 \section{Consequences and Discussion}
We have demonstrated that for a large class of models  in excellent agreement with the current concordance cosmology inflation is naturally followed by an oscillon-dominated phase, provided that the couplings to other fields are small.   These oscillons are generated by parametric resonance,
which occurs if the inflationary potential turns over from the slow-roll regime to a quadratic regime at a scale $M\ll\mpl$.

The inflationary models here are self-resonant, so oscillon production does not require specific couplings to other fields.  It is likely that any significant   couplings between the inflaton and other fields  can
 inhibit  the formation of oscillons, by allowing resonant production of quanta of these additional fields. Further, couplings to other fields can reduce the stability of oscillons by providing an additional channel into which they can radiate energy.  Lastly, the impact of  interactions {\em between} oscillons  is largely unexplored (however, see \cite{Hindmarsh:2007jb}).

Many resonant models include light fields, leaving the universe in an intermediate state between matter and radiation \cite{Podolsky:2005bw,Dufaux:2006ee}, but massive self-resonant models lead to an oscillon-dominated universe that is effectively matter dominated. Our simulations do not include local gravity, but perturbations with sub-horizon wavelengths will grow gravitationally during the oscillon-dominated phase.    The same behavior is seen in non-resonant models with an (almost) homogeneous inflaton condensate oscillating in a pure $m^2\phi^2$ potential \cite{Easther:2010mr}. However, in this case the primordial density fluctuations are $\mathcal{O}(10^{-5})$ at the scale of the horizon and  take a long time to become nonlinear. By contrast, fluctuations grow rapidly in a self-resonant model, leading to a significant enhancement in the primordial power spectrum for high $k$.   The possibility of gravitational collapse and even primordial black hole formation during this phase must be carefully analyzed \cite{1985MNRAS.215..575K, Anantua:2008am}. Given that the oscillons exist on comoving scales vastly shorter than those which contribute to large-scale structure formation, oscillon formation is unlikely to directly modify the primordial power spectrum on present-day astrophysical  scales. However, for any  inflationary model the observed power spectrum is  a function of the post-inflationary expansion history \cite{Adshead:2010mc,Mortonson:2010er}. Thus, it will be important to account for the existence and duration of any matter-dominated phase, oscillon-dominated or otherwise, when computing the detailed predictions of the model.

In summary, we have  shown that a significant class of  realistic inflationary models can naturally lead to 
 copious oscillon production following inflation, and that these oscillons can -- for a time -- dominate the overall matter-density of the universe. This  provides a dramatic example of the potential importance of nonlinear dynamics in scalar fields   to the properties of the very early universe.  

 \section*{ Acknowledgments:}  We  thank Alan Guth, David Shirokoff, Ruben Rosales and Evangelos Sfakianakis for useful conversations. RE and RF are partially supported by the Department of Energy (DE-FG02-92ER-40704) and  NSF (CAREER-PHY-0747868). MA is supported by a Pappalardo Fellowship at the Massachusetts Institute of Technology. HF is supported, in part, by the United States Department of Energy Computational Science Graduate Fellowship, provided under grant DE-FG02-97ER25308. MH is supported by NSF grant PHY-0756174 and a Kavli Fellowship. This work made use of the facilities and staff of the Yale University High Performance Computing Center.

\end{document}